# Examining norms and social expectations surrounding exclusive breastfeeding: Evidence from Mali

Cristina Bicchieri
Professor, Center for Social Norms and Behavioral Dynamics,
University of Pennsylvania
(cb36@sas.upenn.edu)

Upasak Das
Presidential Fellow, University of Manchester
Affiliate, Centre for Social Norms and Behavioral Dynamics, University of Pennsylvania
(upasak.das@manchester.ac.uk)

Samuel Gant
MPP candidate, Harvard Kennedy School
MBA candidate, Tuck School of Business
(samuel_gant@hks.harvard.edu)

Rachel Sander
Technical Research Analyst, MDRC
(rachel.l.sander@gmail.com)

**Abstract**

*Why do people engage in certain behavior? What are the effects of social expectations and perceptions of community behavior and beliefs on one's own behavior? Given that proper infant feeding practices are observable and have significant health impacts, we explore the relevance of these questions in the context of exclusive infant breastfeeding behavior using social norms theory. We make use of a primary survey of 976 mothers of children below the age of two years conducted in 2019 in the Kayes and Sikasso region of Mali, which have a historically lower prevalence of exclusive breastfeeding. Among other information, we collected self-reported data on women's social expectations about the beliefs and behaviors of other community members. The findings from regression estimations, controlling for a host of potential confounding factors, indicate that expectations about the behavior of other community members can strongly predict individual exclusive breastfeeding. Beliefs about community's approval of the infant feeding behavior though are found to be only modestly associated with it. In addition, mothers who hold false but positive beliefs about the community are found to exclusively breastfeed their kids. Further, using responses from randomly assigned vignettes where we experimentally manipulated the levels of social expectations, our data reveal a strong relationship between perceived prevalence of community level exclusive breastfeeding and individual behavior. This result indicates the existence of a potential causal relationship. We argue that our findings represent an important foundation for the design of policy interventions aimed at altering social expectations, and thus effecting a measurable change in individual behaviors. This type of intervention, by using social norm messaging to end negative behavior, avoids the use of coercive measures to effect behavior change in a cost-effective and efficient way.*

**Acknowledgements:**

*We acknowledge the support from the Malien Comité National d'Ethique Pour la Santé et les Sciences de la Vie. We received extensive support and guidance from the Center for Social Norms and Behavioral Dynamics of the University of Pennsylvania (normsandbehavior.sas.upenn.edu)), which received the research grant. We would like to thank Ndiaga Seck and Boureima Konate of the UNICEF Mali Communication for Development section, and Mouhamadou Gueye and his colleagues at CERIPS for their support during the data collection process. We are grateful to all the respondents in Sikasso and Kayes who gave responses to our questionnaire.*

**Introduction**

      Policy makers and academics alike often grapple with the following question: why do people do what they do? This question is especially salient in the context of collective behaviors, which may be deeply embedded in community discourse and are transmitted inter-generationally. Attempts to change collective behaviors by shifting factual beliefs often yield lackluster and impermanent results, indicating that behavior is conditional on other elements (Petit and Zalk, 2019). In contrast, collective behavior change interventions are often known to accelerate social transformation at community level with significant behavior changes at the family, peer group, and individual level. Accordingly, using social norms to alter collective behavior has emerged as a dominant strategy, especially when there is a need to motivate others to act pro-socially (Steg & Vlek, 2009). As argued by Mackie et al. (2015), positive behavioral practices like avoiding child marriage can be encouraged through addressing social motivations, which take into account the broader community's beliefs and actions. Studies have indicated these social motivations can explain the prevalence of certain behaviors, such as open defecation, child marriage, corruption or low female labor force participation (Bicchieri et al. 2014; Jayachandran, 2019; Bursztyn et al. 2020; Bicchieri et al. 2020)

      Using primary survey data conducted in 2018 collected from the mothers of 976 infants under the age of two years from different regions of Mali, we diagnose the collective behavior of exclusive infant breastfeeding (children under 6 months of age) to understand whether it is motivated by the mothers' social expectations. In particular, we use information about individual perceptions of community behavior and beliefs to assess how these perceptions influence the breastfeeding behavior of women with children of two years or younger. To gauge this, we use primary survey data that surveys the respondents about their social expectations and then experimentally manipulate their levels through randomly administered vignettes. This allows us to establish the existence of a causal link between perceived community behavior and beliefs and the practice of exclusive breastfeeding. Our research is especially important in the context of Mali, where studies have indicated while the practice of breastfeeding children is ubiquitous, but the prevalence of *exclusive* breastfeeding of children under six months remains very low (UNICEF, 2017).

      Early exclusive breastfeeding is important as appropriate child feeding practices are vital for the survival and health of children. The World Health Organization (WHO) recommends that

children younger than six months be exclusively breastfed in order to expose them to disease-fighting antibodies in colostrum and to protect them from contaminated water. Studies have found evidence of the significant impact of early initiation of breastfeeding on reduction of overall neonatal mortality (Edmond et al, 2006). In the first six months, it has been found that breast milk alone can provide all necessary nourishment and also carries antibodies that protect the children from a number of ailments including diarrhea and acute respiratory infections. In addition, breastfeeding stimulates the immune system and cognitive development (Butte et al. 2002; Mortensen et al. 2002; Dorea, 2009; Cai et al. 2012; Demir et al. 2020). Notably, it has been estimated that close to 1.4 million deaths of children can be prevented annually in low income countries with optimal breastfeeding for children (Jones et al. 2003).

In this paper we use Social Norms Theory for understanding and diagnosing the collective behavior associated with the self-reported child feeding behavior. Broadly, as argued by Bicchieri (2006) and Bicchieri (2016), the preference for engaging in independent behaviors is not motivated by what others do or approve/disapprove of. In contrast, interdependent behaviors are conditional upon expectations that take the form of Empirical Expectations (EE) (one's expectations regarding what others do) or Normative Expectations (NE) (one's expectations regarding what others think one should do). Collective behaviors that are conditional on empirical expectations alone are *descriptive* norms, while collective behaviors that are conditional on both empirical and normative expectations are *social* norms. This paper examines whether variation in these expectations can alter behavior pertaining to exclusive infant breastfeeding in Mali.

Using a simple probit regression framework, we first assess the relationship between the exclusive breastfeeding of the mothers and their empirical and normative expectations, controlling for other possible individual and household level confounders. Notably, to account for the potential Omitted Variable Bias (OVB) that may restrict us to view the regression results from a causal lens, we add community dummies to control for actual exclusive breastfeeding behavior at the community level along with heterogeneities that include effects from awareness levels, interventions and local institutions among many others. Further we also examine the effect of potential false beliefs on the probability of exclusive breastfeeding by regressing the bias in perception on the behavior, controlling for the other confounding factors. Even after accounting for the unobservable, mothers who breastfeed their children might be more likely to hold false beliefs about the community; we thus present vignettes with random combination of four mutually exclusive types of expectations: low EE-low NE; low EE-high NE; high EE-low NE; high EE-high NE (Bicchieri et al. 2014). Notably, these vignettes are assigned randomly to the respondents, who are asked to respond about the likelihood of breastfeeding behavior in the context of the information received. Because these vignettes are experimentally manipulated and randomly administered, we argue that the inferences from our empirical design account for OVB as well as reverse causation, which allows us to give a causal interpretation to the relationship between expectations and behavior.

The findings from the regressions indicate a significant and positive relationship of EE and also NE with exclusive breastfeeding behavior, though the effect of EE is found to be considerably stronger. The results remain robust across different specifications and also after adding community level dummies that control for the actual prevalence of breastfeeding in the community. We also find false beliefs to be significantly related to behavior: respondents who have false but positive beliefs about the prevalence of exclusive breastfeeding in the community are more likely to exclusively breastfeed and those with negative false beliefs are less likely to do so. Importantly, through randomly assigned hypothetical vignettes, we find respondents who are presented with the high EE and high NE combination have significantly higher chances of reporting exclusive breastfeeding as their own likely behavior. Interestingly, for respondents who are given the high EE-low NE combination, the chances remain similar to the previous case. The responses from the low EE-high NE type vignette show a massive drop in predicted breastfeeding, indicating a robust positive effect of EE on exclusive breastfeeding behavior while the effect of NE is modest. This research paves the way for innovative community level interventions that can alter existing social norms and go a long way in improving exclusive breastfeeding among women in Mali and other countries with similar context and beliefs.

The paper has multiple contributions. First, we contribute to the studies on the highly prevalent problem of non-exclusive breastfeeding in the developing world in general and in Mali in particular. These findings may be relevant in other African countries with similar socio-economic parameters and cultural beliefs. Second, we contribute to the emerging literature on issues related to social norms, gender and public health. Here, we use the SNT framework and assess whether changes in social expectations and beliefs can lead to positive changes in personal behavior. We utilize the findings to catalyze a discussion on how interventions to shift social expectations can cause a measurable change in individual behaviors. Demonstrating a causal relationship between social expectations and individual behaviors provides a foundation for the design of interventions that can sustainably change collective behaviors by shifting social expectations. This constitutes the main contribution of the paper.

The structure of the paper is as follows. Section 2 discusses the framework of the social norms theory that we use to examine the changes in exclusive breastfeeding behavior. Section 3 discusses the data and variables while section 4 presents the empirical strategy. Section 5 discusses the findings, and the relevant policy implications are discussed in section 6. Section 7 presents a brief conclusion.

## 2. Social Norms Theory

The study of social norms begins with the study of collective behaviors or behaviors that are common to many people. The distinction between social norms and other collective behaviors lie in the motivation that drives the behavior. It is generally agreed upon that social norms are driven by the beliefs and/or practices of other people belonging to one's community, or at least among people whose opinions matter with respect to certain decisions. In other words, people follow social norms in order to avoid social sanctions which can take the form of ostracization, isolation, boycotts or even just being frowned at. For example, many people feel obliged to drop change in a coffee shop tip jar because they don't want the baristas to think they are stingy. Capitalizing on this fear of shame, coffee shop workers may "salt" the tip jar. This serves as an indication that other customers left tips, encouraging the next customer to leave one too.

Social norms may have positive or negative impacts on society. For example, when individuals throw away litter in trash cans in public, others may imitate them and society at large benefits. On the other hand, studies have indicated norms of paying bribes to police in Nigeria benefit the police, but are damaging to the larger community and negatively impact social trust (Patel and Hoffman, 2017). Similarly, widely shared norms are often inefficient and even harmful, as when parents choose to arrange a daughter's marriage while she is still a child. They may act out of fear that the girl will get pregnant out of wedlock, bringing shame to the family. The social norm of child marriage persists in many regions around the world despite the clear relationship between early marriage and poor educational outcomes, a higher risk of health problems and financial dependence.

There are various theories on how to specifically define a social norm. Some define social norms as any behavior that one follows in order to blend in with others in their network (Burke and Young, 2011). When someone witnesses a behavioral pattern in their community, they want to adopt the behavior in order to be accepted. Here, the defining feature of a social norm is its reinforcing nature: the more widespread the norm among a particular group is, the more other people in that group want to follow it. One example is following a fashion trend. As more people adopt the fashion, others will join in to be included in the group.

Norms, however, are often implicit and informal, and are maintained through tacit beliefs and expectations. People follow a certain behavioral pattern, not only because of what they believe to be commonly practiced, but also because what they believe others approve of and expect them to do. From this perspective, a society does not need to discuss openly what should be done in order to preserve the social norm. People's behavior is based on their beliefs about what others do and/or approve of, even if those beliefs are false. (Bicchieri 2006; Reid et al., 2010; Siu et al., 2012; Sunstein, 1996; Vaitla et al., 2017).

Here it is important to distinguish between *descriptive* and *social* norms (Bicchieri 2006) Descriptive norms, such as fashions, fads, rules of etiquette and conventions (from language to traffic rules), are followed because one prefers to imitate or coordinate with common behaviors that are expected to occur. If the goal is to coordinate with others, for example, it is perfectly rational to follow common rules (e.g. obeying traffic signals). Social norms are more complex because they usually exist to reconcile individual and common welfare, which may differ.

As an example, after eating in a restaurant there is no punishment for not leaving a tip, and all individuals would prefer not to pay more than necessary for their meal. However, if an individual believes that other patrons will leave tips and they think that others who matter to them would disapprove of not leaving a tip, an individual is more likely to leave a gratuity. With descriptive norms, empirical expectations (i.e. expectations about what others do) are sufficient to motivate the behavior. With social norms, compliance with the behavior depends on one's empirical expectations *and* normative expectations (i.e. expectations about what others think one should do To extend the previous example, an individual who is accustomed to tipping in restaurants in her own country may not tip when she travels to another country, where few other patrons leave tips and there is not a social expectation that one should tip. Thus the decision to tip is not absolute, but rather preferential: an individual prefers to tip when certain empirical and normative expectations are met (Bicchieri 2006). The consequences of deviating from the norm account for another important element in social norms theory. Unlike legal sanctions, which are imposed by law enforcement, social sanctions are enforced by community members (Miguel and Guherty, 2005). Despite the fact that these punishments are almost never formal, they are often more effective than legal sanctions.

From a behavior-change perspective, defining norms in terms of expectations and preferences has several advantages. First and foremost, we can measure these expectations. By asking targeted questions about empirical and normative expectations, we can understand the degree to which one's perceptions about others influence behavior. Moreover, if we find that behavior is conditional on social expectations, we know where to act. If behavior is conditional on expectations, a shift in expectations will cause a shift in behavior. Public health initiatives with the goal of ending harmful practices often target individuals' attitudes, but these are neither strong predictors of behavior nor do they address pressure from the community, whether real or imagined. Imagine a group in which most individuals do not endorse child marriage but believe most people in their group support it. If one fears that openly criticizing this practice would put one at a disadvantage, child marriage will survive. In this case, a norm may be followed, but not be personally endorsed by a majority of individuals. The discrepancy between "perceived norms" and "actual norms" suggests that people may have incorrect perceptions about what others believe is right and wrong. In this case, revealing the true majority beliefs may change a longstanding norm.

In this study, we use the Social Norms Theory (SNT) framework to understand and diagnose collective behaviors in the context of exclusive breastfeeding (Bicchieri 2006; 2016). SNT identifies four broad categories of behavior which have been outlined in Figure 1 and are described below: all collective behaviors are either socially independent or interdependent. Independent behaviors refer to the behaviors that people adopt irrespective of their beliefs regarding the actions and opinions of others. Interdependent behaviors on the other hand are motivated by social expectations (what we believe others do and think). Two examples of independent behaviors are customs and moral rules. Customs are driven by practical needs whereas moral rules are driven by values. For example, when it rains, many people use an umbrella. Their decision to do so is independent of other people's decisions to use an umbrella. It simply serves a practical need to avoid getting soaked. The use of umbrellas is a custom: a convenient action that satisfies a need. Another example of independent behaviors are the rituals followed by many religious people. These practices are observed independently of others' behaviors, and are motivated by convictions about what is "right" and "wrong" rather than by pragmatism. For example, individuals with strong beliefs on some particular religion may always avoid particular food items, even if those items are more affordable than others or even when other people around them are eating those food items. A moral rule is neither utilitarian nor is it dependent on others' actions. Instead, personal normative beliefs about what a "good" individual who subscribes to that religion should do, drive this behavior.

Contrary to independent behaviors, interdependent behaviors include descriptive and social norms. As outlined above, interdependent behaviors are based on social expectations: empirical and normative. Notably, the term "expectations" is used instead of "observations" because one is not always correct in the estimations of what others are doing or thinking (Bicchieri 2006, 2016).[1]

Figure 1: Categories of behavior under SNT

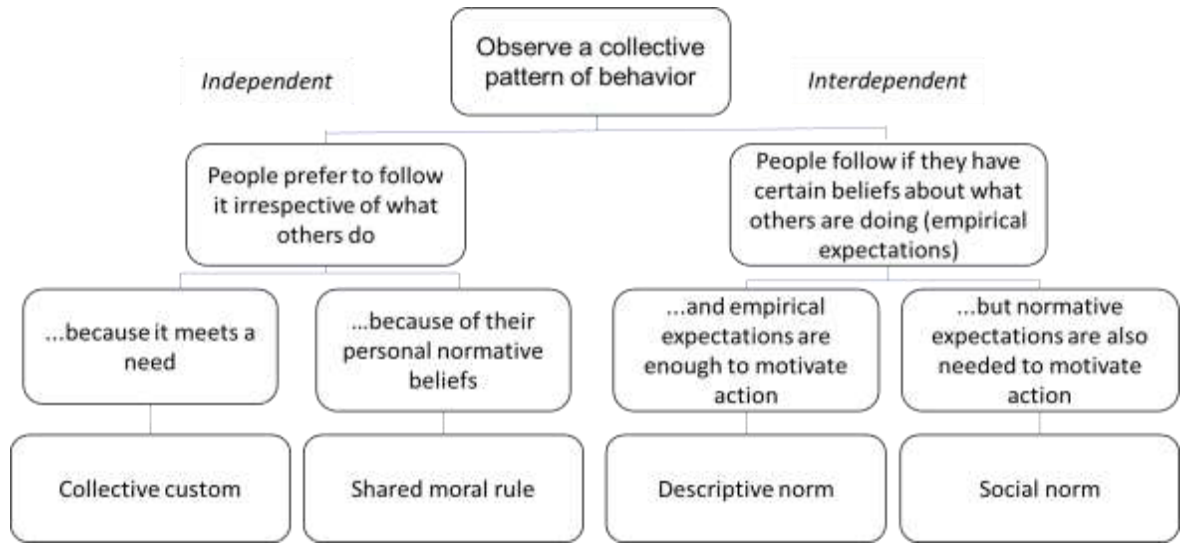

---

[1] Bicchieri (2006, 2016) carry many examples of these different types of behaviors.

This paper applies the SNT framework and studies the implications of empirical and normative expectations on exclusive infant breastfeeding behavior in the Malian context. In particular, we examine if there is a descriptive norm or social norm associated with this behavior. Important to note here is the fact that exclusive breastfeeding behavior is often publicly observable.[2] In the Malian context, mothers can usually observe the infant feeding behavior of their peers, and thus can be influenced by what they see. Dettwyler (1987) notes that there is no taboo around public breastfeeding and observes "..women grow up expecting to have children as a normal part of their adult lives. They have ample opportunity to observe their own mothers, other women in the compound, and other women in the community progressing through many pregnancies and nursing the babies after birth." (pp. 635). Other studies have indicated that social expectations may drive behavior such as exclusive breastfeeding among mothers in a community (Mickey et al 2015). Of note is the fact that the average birth rate in Mali is close to six children for every woman. When combined with the early age of marriages, this would indicate that many Malian women will be pregnant or nursing for most of their first ten years of marriage.

**Data**

The data used in the study has been collected through a primary survey conducted in March, 2019 in the Kayes and Sikasso region of Mali. The objective of the project is to better understand the psychosocial factors that hinder behavior change with respect to non-exclusive breastfeeding and complementary feeding (Research protocol approval number: 0022/MSHP/CNESS). The study participants were identified using data from the country's most recent General Population and Housing Census (RGPH 2009), administered every ten years by the Malian National Institute of Statistics (INSTAT). INSTAT draws the boundaries of Mali's enumeration sections (ES), which are areas containing roughly 100 to 200 households. To reflect the larger proportion of rural SEs over urban ESs in Mali (in the two regions under study, 15% of the population is urban and 85% of the population is rural), the sample was stratified by region (Kayes vs. Sikasso) and by urbanization (urban vs. rural). Urban areas were over-sampled, allowing the study to derive meaningful information for what would otherwise be an insignificant sample size of urban households. In each region, 40 ESs were randomly selected: 70% (28) were rural and 30% (12) were urban.

---

[2] To assert that certain behavior is informed by social expectations, it is necessarily not the case that the behavior (or its consequences) needs to be publicly observable. As an instance, even 'private' behavior like premarital sexual engagement is often informed by social expectations.

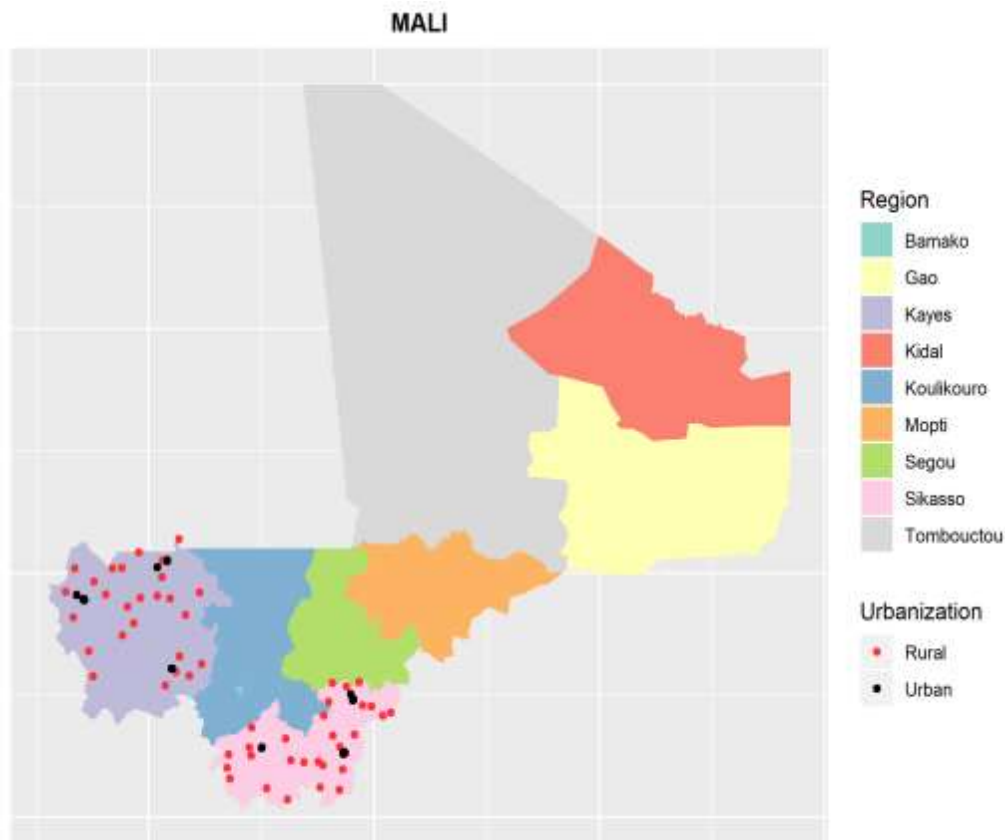

Distribution of 80 enumeration sections in Mali

Within each of the 80 selected ESs, surveyors conducted a rapid enumeration of all households to determine which were eligible for participation in the survey. Eligibility meant that at least one child (under 18) or pregnant woman lived in the home. Following the enumeration, 25 eligible households were randomly selected in each ES to participate in the survey. Within each household, researchers endeavored to survey the mother of the child living there, the head of household (usually the mother's husband), and, if present, a third generation woman (usually the mother of the head of household). The study's eligibility criteria did not include the presence of this third generation woman, in order to capture data from a broader selection of family structures. As a result, in nearly two thirds of the targeted households, no third generation woman was available, primarily because she had passed away, was hard of hearing, or wasn't well enough to participate in the survey. Only 1.6% of sample respondents who were located in the field refused to consent to the survey.

Overall, 3,956 people were surveyed of which 2,535 were female. These women include 11 heads of household, 1,850 mothers of children living in the home, and 674 mothers of heads of household. All female respondents were asked the ages of their living children. If a woman reported that she has a child younger than 2 years, she was asked to answer questions on child feeding practices. Of the 2,535 female respondents, 761 were mothers of children 2 years or younger. These women account for the 976 infants who make up the research sample.

The survey was written in French but was administered in local languages, primarily Bambara (93%) and Soninke (5%). Extensive back-translations were conducted to ensure the accuracy of translations. The survey tool was tested in a three-day field pilot that realistically reproduced the enumeration, informed consent, and interview process. A structured questionnaire to gather data that would help diagnose the target behaviors was developed. It comprised of the following modules:

**Demographics:** The survey measured each respondent's age, sex, marital status, ethnic group, household size, highest level of education reached, floor and roof materials, estimated monthly income, and religion. Questions to gather information on these were asked directly to the respondent.

**Self-reported behavior:** With respect to child feeding, the survey asked all mothers with children two years or younger to report about the food products each child was fed from 0 to 6 months and from 6 to 24 months (if the child is currently more than 6 months.) These mothers were also asked how soon after giving birth did the child receive breast-milk.

**Empirical expectations:** To get an understanding of the empirical expectations and quantify them, surveyors asked "*Out of 10 mothers in your community, how many do you think exclusively breastfeed their babies during the first six months*." The question was preceded by a "statement of normalization" (i.e. "*In some communities, mothers exclusively feed their babies breastmilk during the first six months. In other communities, mothers do not do this*.") This statement intends to reduce social-desirability bias. For this question, the answering options were coded as:

0) None [0]

1) Few people [1-4]

2) Half of the people [5]

3) Most people [6-9]

4) Everyone [10]

5) I don't know

**Normative Expectations:** Using the same structure as empirical expectations, surveyors asked each respondent their expectations about what others *think they should do*. In particular, the surveyors posed the questions: "*Out of 10 people in your community, how many believe that a mother should not give 'decoctions' to her child before 6 months?*" The answering options on this question were coded as:

0) Nobody [0]

1) Few people [1-4]

2) Half of the people [5]

3) Most people [6-9]

4) Everyone [10]

5) I don't know

We use this question to understand about the respondent's perception on whether the community approves or disapproves of exclusive breastfeeding. In other words, it can be assumed that if an individual believes a mother should not give decoctions to her child in the first 6 months, she/he approves of exclusive breastfeeding and vice versa.

**Vignettes:** We used vignettes to assess the conditionality of exclusive breastfeeding and to understand the causal influence of empirical and normative expectations. In a short story about an imaginary character, we experimentally manipulate the level of empirical and normative expectations and measure their effect on predicted behavior. These vignettes had been administered in a 2x2 framework in which each respondent was read a random vignette including high or low empirical expectation and high or low normative expectation. Given these parameters, the respondent then predicts the character's behavior. This allows us to gauge the extent to which a change in EE or NE leads to changes in the respondents' behavior. Through the vignettes, we measure the respondent's assessed likelihood that the vignette character will behave in a specific way. One assumption in our setting is this would also indicate how the respondents themselves would respond to exogenous random changes in expectations. Important to note here is that the vignette protagonist should be a fictitious character which ensures that it is easier for respondents to have hypothetical thinking and minimize social desirability bias. In our survey, the following question was asked:

> *A young woman named Mariam, who you don't know, moved from your area to this community, one year ago. Mariam has just given birth to her first child. In this community [most]/[few] mothers exclusively breastfeed their children up until 6 months old. Also, in this community, [most]/[few] people believe that a good mother should exclusively breastfeed her child up until the age of 6 months. What do you think Mariam will do?*

Variables

Our outcome variable is derived from the self-reported behavior on exclusive breastfeeding. We code "1" if the mother reported exclusive breastfeeding and "0" otherwise. The two main variables of interests are derived from the questions posed on EE and NE. Both variables EE and NE are coded as "1" if the respondent answered "most people" or "everyone" and they are coded as "0" if the respondent answered "nobody", "few people" or "half of people. Therefore this coding tells us if the respondent believes that a majority of the individuals in the community exclusively breastfeed the infants or not, and also if the respondent believes the majority approves of the behavior or not. As one can see, the main variables of interest as well as the outcome variable in our case are binary in nature.

A battery of control variables that can potentially confound the estimates about actual behavior have been incorporated in the model. Since the mother's age is a determinant of child breastfeeding in general, we include it as a control variable (Jayachandran and Kuziemko, 2011). Education is also a likely confounder as it has a direct potential bearing on breastfeeding and hence is included in the model. We also include whether the respective household is polygamous. The economic status of the household is controlled through whether the respondent has a "good" house identified through indicators like a cement roof and floor. In rural areas, a "good" house also includes homes with metal roofs and stone floors. In addition we add controls for whether or not a household received any remittance, which also controls for the fact that the household includes a migrant. Further, we add a variable that identifies whether the respondent has been educated on health related issues by any Non-Governmental Organization (NGO) since that may have a positive relationship with breastfeeding behavior. To capture the region specific and ethnic heterogeneities that may affect the outcome variable, we add indicators for the "cercle", which are the second largest administrative level in Mali after the region. This is our preferred specification. However in the last specification, we also include ES fixed effects in order to control for the community level heterogeneous variations to check if there is a substantial deviation in the estimates qualitatively as well as quantitatively. The standard errors are clustered at the ES level to account for within community variations.

**Empirical Strategy**

To estimate the effect of empirical and normative expectations surrounding breastfeeding on the actual behavior, we first make use of a simple probit regression, which yields the causal estimates through maximum likelihood. In particular, we estimate the following regression equation:

$$Pr\ (Y_{ihc} = 1)\ = F(\beta_0 + \beta_1 EE_{ihc} + \beta_2 NE_{ihc} + \beta_3 I_i + \beta_4 H_h + \beta_5 C_c \qquad (1)$$

Here $Y_{ihc}$ is the outcome variable which takes the value of 1 if mother $i$ from household $h$ and cercle $c$ reports of having exclusively breastfed the child in the first six months, and 0 otherwise.[3] $EE_{ihc}$ denotes the associated empirical expectation which takes the value of 1 if the mother reports that the majority of other mothers within their community exclusively breastfeed their children in the first six months after birth, and 0 for those who think otherwise. $NE_{ihc}$ represents the associated normative expectation which assumes the value of 1 if the mother reports that she thinks the majority of the other mothers within their community approve of exclusively breastfeeding their children, and 0 otherwise. In the above equation, individual level controls are given by $I_i$, household level controls by $H_h$ and dummies associated with cercles as $C_c$. $\beta_1$ and $\beta_2$ are estimated marginal effects of empirical and normative expectations respectively.

---

[3] "Cercle" is a second level administrative division. There are 8 "regions" in Mali, and 49 "cercles" within those regions.

Despite controlling for a host of individual and household level factors along with geographical and community factors through district and cercle fixed effects, it is possible that mothers who exclusively breastfeed their children are more likely to have false beliefs about the community. In fact literature has indicated that an individual's prevalence perception is largely subjective and may not very well reflect the actual frequency of certain behaviors within the community (Suls et. al., 1988; Kruglanski, 1989). Accordingly, we examine whether these perceptions are inaccurate and how this bias is associated with the breastfeeding behavior of the mothers. Therefore in the second set of regressions, the following equation is estimated:

$$Pr(Y_{ihc} = 1) = F(\beta_0 + \beta_1 Bias(EE_{ihc}) + \beta_2 Bias(NE_{ihc}) + \beta_3 I_i + \beta_4 H_h + \beta_5 C_c) \quad (2)$$

Here $Bias(EE_{ihc})$ take the value of "0" if the respondent is able to correctly predict the prevalence of mothers who exclusively breastfeed in the community. So if the respondent thinks few/most mothers in the community exclusively breastfeed their children and our data show that estimate to be the actual prevalence within the respective ES, $Bias(EE_{ihc})$ assumes the value of 0. It takes the value of "1" if the respondent over-predicts the actual prevalence of exclusive breastfeeding within the community and a value of "2" if she under-predicts the same. Similarly, if the respondent believes that few/most mothers in the community approve of exclusively breastfeeding their children under 6 months and we find that to be the actual prevalence within the cercle, $Bias(NE_{ihc})$ takes the value of "0" and then "1" and "2" depending on whether she over-predicts or under-predicts the personal normative beliefs of other mothers within the community respectively.

Existing literature indicates that bias in prevalence perception is positively associated with one's own personal behavior. In particular, it has been found that people tend to perceive their own behavior to be more common in the community than it actually is (Ross et al., 1977; Sherman, 1983). Therefore, we use vignettes to assess the conditionality of exclusive breastfeeding and to understand the extent to which empirical and normative expectations causally determine the behavior. Through hypothetical vignettes, we experimentally change the levels of empirical and normative expectations through a story and measure their effect on the character's behavior as predicted by the respondent. As discussed, these vignettes have been administered through a 2x2 design, in which each respondent was given a single vignette with random variation in empirical and normative expectations (high or low). Through this, we are able to determine the degree to which exogenous changes in expectations causally determine changes in predicted behavior among our respondents. This serves as an indication as to how the respondent themselves might respond to varying social expectations. Accordingly, we regress the following equation to assess the extent of dependence of social expectations on personal behavior:

$$Pr\ (Y_{ihc} = 1)\ = F(\beta_0 + \beta_1(lowEE_{ihc}lowNE_{ihc}) + \beta_2(lowEE_{ihc}highNE_{ihc}) + \beta_3(highEE_{ihc}lowNE_{ihc}) \\ + \beta_4(highEE_{ihc}highNE_{ihc}) + \beta_5 I_i + \beta_6 H_h + \beta_7 C_c \qquad (2)$$

Here $lowEE_{ihc}lowNE_{ihc}$ denotes the type of vignette that has low EE as well as low NE. Similarly, $lowEE_{ihc}highNE_{ihc}$ denotes that type of vignette, which has low EE but high NE. The others follow the same rule. Note that since the type of vignettes that is given to the respondents is randomly administered, we argue that the answers are solely linked to the conditions that are placed through the vignettes. Therefore, controlling for the potential confounders, estimates ($\beta_1$ to $\beta_4$) can be understood to be causal in nature. These estimates give the impact of the four possible variants of EE and NE together on breastfeeding behavior.

**Results**

Table 1 presents a brief description of the data. In our sample of 976 mothers of children below two years of age, we find that only about 19% were exclusively breastfed before 6 months whereas close to 81% were not. Among these children's mothers, close to 23% believe that the majority of their community exclusively breastfeed their kids in the first six months and about 30% believe that the majority also approve of the exclusive breastfeeding behavior. Importantly, we observe more than 55 percent of the women who breastfeed their children exclusively in the first six months think that the majority of women in their community exclusively breastfeed.

About 55% of them also think that majority in the community approve of exclusive breastfeeding in the first six months.

[Table 1 here]

*Basic regression*

Table 2 gives the estimation results from the regression outlined in equation 1. We present five specifications: the first has indicators on empirical and normative expectations in the model; the second includes the individual and household level controls, the third introduces the cercle fixed effect variables, the fourth incorporates the fixed effects on ethnicity and the final specification includes the ES dummies along with all the other variables. This final specification has the most comprehensive set of controls that allows us to get close to unbiased estimates and hence it is the one which we prefer to use in the next set of regressions. The findings from the regressions indicate that the probability of a mother exclusively breastfeeding her child increases by more than 22 percentage points for women who think that the majority of other mothers in the community exclusively breastfeed. We also find a 10 percentage points increase in the likelihood for women who think that the majority of other mothers in the community approve of the exclusive breastfeeding behavior. This underscores the importance of social expectations and their possible influence on exclusive infant breastfeeding behavior.

In terms of the control variables, the individual and household level indicators are found to be statistically insignificant though the women from Sikasso are less likely to exclusively breastfeed their children. This may be attributable to what has been referred to as the "Sikasso Paradox." Sikasso has the most rainfall and greatest level of crop diversity of any region in Mali, but counter-intuitively inhabitants of the area are disproportionately poor and malnourished. A 2013 assessment by the World Bank found that although only 16% of Malians live in Sikasso, the region contains 34% of Mali's poor population. Sikasso has the highest infant mortality rate and the highest under-five mortality rate in the country (Cooper and West, 2017). A higher regional rate of non-exclusive infant breastfeeding may be a partial contributor to these poor health indicators. It should be noted that when ES dummies are included, some of them get dropped because in those ES, the outcomes get predicted perfectly. To ensure there is no bias in the estimates because of the drop in the sample, we use inverse probability weighing in the regressions, which can be used in case of high attrition or missing observations (Wooldridge, 2007). The results are found to be robust to this technique as well.[4]

[Table 2 here]

*Relationship with perception bias*

As indicated earlier, it is possible that individuals may have false beliefs about the community and hence we examine if false beliefs are in any way related to behavior. Figure 1 presents the regression estimations which show how positive and negative bias regarding beliefs about community level breastfeeding norms are associated with personal behavior. Interestingly, we find that women who underestimate exclusive breastfeeding in their community are less likely to engage in the exclusive breastfeeding behavior. There is a close to 23 percentage point jump in the probability of exclusive breastfeeding for a woman who overestimates the number of women in the community exclusively breastfeed their newborns for the first six months. This is with respect to those mothers, who correctly estimate the exclusive breastfeeding rates in their community.

[Figure 1 here]

In comparison, the bias related to normative expectations are found to have a weaker association with behavior. Controlling for individual and household characteristics as well as community level heterogeneities, there is a 15 percent chance that a woman who underestimates women's approval of exclusive breastfeeding in the first six months will exclusively breastfeed her child in the first six months. However, unlike the earlier model where we found a substantial jump in the probability of exclusive breastfeeding among women with positive bias in empirical expectation, the increase here is a relatively meager 10 percentage points (though statistically significant at 5% level).[5]

In other words, our findings show that women who overestimate the prevalence of exclusive breastfeeding in the community are substantially more likely to exclusively breastfeed. We also found that women who overestimate the approval of exclusive breastfeeding within the community are more likely to exclusively breastfeed in the first six months. However the effect size of the former is found to be much higher than that for the latter. This finding coupled with the finding from the normal probit regressions indicates empirical expectations are more likely to predict exclusive breastfeeding behavior though normative expectations also have a modest role to play.

---
[4] The results from inverse probability weighing can be provided on request
[5] The marginal effects from the regressions are shown in Appendix A1.

*Estimations from randomly assigned vignettes*

As indicated earlier, despite controlling for the potential confounders in the form of the individual, household, geographical and community factors, it is possible that mothers who exclusively breastfeed their children are more likely to believe that other mothers in her community share their beliefs about her community in terms of breastfeeding Therefore, to argue that social expectations determine breastfeeding behavior, we present vignettes, each with one of four possible combinations of high/low EE and NE. So we have four possible vignettes, one of which is randomly assigned to each respondent: low EE and low NE; low EE and high NE; high EE and low NE; high EE and high NE. We then examine the answer corresponding to the given vignette. Because of this random exogenous variation of the vignette presented to the respondent, we argue that the marginal effect associated with each combination gives us the causal estimate of social expectations.

Before we present the regression results, we establish that the combinations of vignettes that are presented to the respondents are random in nature. To gauge this, we show the prevalence of exclusive breastfeeding behavior separately for women presented with these four different combinations. Figure 2 shows the proportion along with 95% Confidence Intervals (CI) calculated through the standard errors clustered at the ES level. We find no significant difference in the prevalence of exclusive breastfeeding among women in these four categories as the CIs are found to overlap one another.

[Figure 2 here]

Next, we group the respondents based on the vignette they were presented and we compare the mean values of the control variables used in the regression across these groups (Table 3). The findings reveal no substantial difference across the four groups. For example, the average age of women is 26 to 27 years across all four groups. Similarly, around 30% of women practice polygamy and this is the case across all the four groups. However, for certain variables, we do find some difference, but these differences are not statistically significant at 95% level. Further even if one argues that this imbalance biases the causal estimates, our regression strategy controls for these confounders and hence we argue that the regression estimates are unbiased.

[Table 3 here]

Figure 3 presents the marginal effects from the main regressions with vignettes. As one can observe, when the respondent is presented with a vignette with low EE and low NE, the chances of exclusive breastfeeding is found to be close to 40%. When the vignette changes to a low EE with high NE, the chances increase by close to 9 percentage points on average. However as we move from low EE to high EE, we see a jump in the likelihood that the respondent predicts the subject of the vignette will exclusively breastfeed. This probability is as high as 78% for high EE and low NE vignette and 80% for high EE and high NE vignette. Of note is the fact that we obtain these findings after adjusting for individual, household and community factors as well as actual behavior (e.g. whether the respondent exclusively breastfeeds). The findings clearly indicate a causal influence of social expectation on breastfeeding. Empirical expectations have a stronger causal influence, and hence breastfeeding behavior can be said to be a descriptive norm rather than a social norm. Importantly, these inferences validate the results obtained from laboratory experiments, where it is found that EE supersede NE especially when they are incongruent (Bicchieri and Xiao, 2009)

Notably these findings hold across different specifications, which tend to indicate that the inferences are robust.[6] This underscores the importance of empirical expectations in inducing the socially desirable behaviors. However, this analysis through vignettes gives the likelihood of possible exclusive breastfeeding behavior of the respondents and does not take into account the actual change in behavior in response to real changes in empirical and normative expectations. In future research one can do interventions that randomly feed information aimed to alter social expectations and examine its influence on actual breastfeeding behavior among mothers.[7]

---

[6] The marginal effects from the regressions with different specifications are shown in Appendix A2.
[7] Notably, Bursztyn et al. (2020) assessed how providing information about other men's perceptions on women participating in the labor market impacts their own approval and willingness to let women from their households to join the labor force.

[Figure 3 here]

In terms of other potential confounders to the vignette response, as one would expect, the women who exclusively breastfeed are significantly more likely to report that the subject of the vignette would exclusively breastfeed, even after adding other controls including the type of vignette presented to the respondent. With respect to the effect size, if the respondent exclusively breastfeeds her infant, she is more likely to predict that the subject of the vignette will exclusively breastfeed by about 34 percentage points. Education also seems to have a significant effect as we find that educated women are more likely to predict that the subject of the vignette will exclusively breastfeed in comparison to illiterate women by about 10 percentage points.

**Policy Recommendations**

Our main finding from the paper is evidence of a significant positive influence of increasing EE of exclusive breastfeeding behavior, while we observe modest effects of a change in NE. This means that perceptions of common behavior are more critical to behavior change than perceptions about the community's approval of the behavior. These insights can especially help in designing policy instruments that encourage a behavioral change towards exclusive breastfeeding. The role of social expectations in driving exclusive breastfeeding should be acknowledged by policy makers and implementers. Behavior change interventions should aim at altering the existing beliefs and biases with regard to breastfeeding. Since we diagnose exclusive breastfeeding in the Malian context as mainly a descriptive norm, that is, it is substantially driven by EE but only moderately by NE, we conclude that a change in the former may prove to be a necessary precursor to a change in behavior. Policy makers can shape/change social expectations in order to effect a change in the behaviors that depend on those expectations.

One method that has been used to change descriptive norms is known as *descriptive norm messaging*. For example, in India open defecation remains a threat to public health and has been diagnosed by researchers as a descriptive norm. That is, the practice remains widespread in certain areas because those community members believe it is prevalent among their neighbors, which drives a belief that there is sufficient acceptance to continue the practice. Researchers are targeting these empirical expectations by disseminating messages that highlight the emergence of a new norm (exclusive toilet use) and the decline of the old norm (open defecation) (Bicchieri et al. 2018; Ashraf et al. 2020). The recipient of this message begins to view open defecation as a deviation from the norm, and, in the case of descriptive norms, will abandon open defecation in favor of the more widely practiced behavior. In a similar vein, messaging in the breastfeeding case can use the positive impact of EE to effect collective behavior change. This could be done by sending messages like "*many of the mothers within the community have started exclusively breastfeeding their children or intend to start doing so soo*n." Our research indicates this can potentially increase the likelihood of exclusive breastfeeding. This increase in the prevalence of breastfeeding can then serve as messaging tools to further enhance the EE and as a result improve the exclusive breastfeeding rates. These messages may be delivered by health and NGO workers at community events, by family members within the household, or by one's social network through text messages. An advantage of this approach is that the shift in empirical expectations becomes self-reinforcing as more people adopt the target behavior, making the behavior change more durable. Of note is the fact that strengthening capacity of communities for delivering such messages have been found to be effective in improving breastfeeding practices (Alderman, 2007).

Signalling is another form of descriptive norm messaging, where decals are placed on the front of houses where the desired behavior is adopted. The goal of these decals is to inform people in target communities which of their neighbors already adhere to the desired behavior. The observers' empirical expectations will shift, leading to a potential change in one's own behavior. Notably, such norm-centric behavioral change interventions have been implemented in parts of India to encourage consistent usage of toilets. The idea is to change collective beliefs by altering the beliefs about the community's sanitation practices. Interventions in similar vein may be implemented which can potentially increase the prevalence of exclusive breastfeeding (Ashraf et al., 2021).

**Conclusion**

International development practitioners have increasingly been adopting collective behavioral change models into their Theory of Change. While the literature has various definitions of social norms (Bicchieri et al., 2018), we employ an approach based on beliefs and preferences, whereby behavior is contingent on beliefs about what people in the reference group do and approve of. When there is a need to motivate others to engage in prosocial behavior and the target behavior is driven by social expectations, changing social expectations emerges as a key strategy. Accordingly, in this paper we explore if social expectations can predict and influence exclusive breastfeeding behavior of newborns.

Using self-reported information on the perception of the prevalence of exclusive infant breastfeeding and its approval, the findings reveal a robust, statistically significant association between perceived community behavior and a mother's likelihood of exclusive breastfeeding. Responses from the randomly assigned vignettes with experimentally manipulated levels of empirical and normative expectations reveal these social expectations can potentially change maladaptive collective behaviors like non-exclusive breastfeeding. In particular, we observe that the social expectations of mothers remain among the major contributing factors to non-exclusive breastfeeding during the first six months. This implies that these women believe that most other mothers in the area do not exclusively breastfeed and do not approve of exclusively breastfeeding. Our findings demonstrate the causal influence of empirical expectations on non-exclusive breastfeeding and therefore support the diagnosis of exclusive breastfeeding as a descriptive norm.

Our findings provide a foundation for policy design. There is no plausible injunctive means of commanding and enforcing behavior change in Malian communities. With the newfound understanding that behavior is conditional upon social expectations, it stands to reason that a change in social expectations will be followed by a change in behavior. Descriptive and social norm messaging can shift community members' attitudes by updating and adjusting empirical and normative expectations. Once expectations of the frequency of exclusive breastfeeding start increasing, the desired behavior follows. Given the mutually reinforcing relationship of empirical expectations and actual behavior, a new norm could eventually replace the older one.

Importantly, our observations underscore the importance of initiating an effective campaign to change empirical expectations and assert that this will have significant causal effects on child feeding behavior in Mali. Developing programming of such interventions and evaluation of the impact after implementation remains an area in need of further research. In addition, a question for further study is the prevalence of pluralistic ignorance in specific communities targeted for interventions. In instances where community members consistently underestimate the proportion of their neighbors who breastfeed exclusively (relative to community-side self-reported behavior statistics), there is potential to correct false empirical expectations and rapidly effect social change.

In the study of social norms, accurate, up-to-date, and localized data are essential for correctly diagnosing behaviors and effectively measuring intervention effects. We argue that our findings in Mali may have relevance to similar contexts, where non-exclusive breastfeeding is not prevalent. However, we caution that identical observed behaviors can have different motivations wherein non-exclusive breastfeeding may be conditional on other factors. Therefore, to leverage our findings and to create tangible change in the field, it is necessary to identify target communities, design collective behavior change interventions, and conduct pilots to identify the most contextually specific approach.

Ross, L., Greene, D., & House, P. (1977). The "false consensus effect": An egocentric bias in social perception and attribution processes. *Journal of Experimental Social Psychology, 13*(3), 279-301.

Sherman, S. J., Presson, C. C., Chassin, L., Corty, E., & Olshavsky, R. (1983). The false consensus effect in estimates of smoking prevalence: Underlying mechanisms. *Personality and Social Psychology Bulletin, 9*(2), 197-207.

Steg, L., & Vlek, C. (2009). Encouraging pro-environmental behaviour: An integrative review and research agenda. *Journal of Environmental Psychology*.
https://doi.org/10.1016/j.jenvp.2008.10.004

Suls, J., Wan, C. K., & Sanders, G. S. (1988). False consensus and false uniqueness in estimating the prevalence of health-protective behaviors. *Journal of Applied Social Psychology, 18*(1), 66-79.

UNICEF (2017). Enquête Nationale Nutritionnelle Anthropométrique Et De Mortalité Rétrospective Suivant La Méthodologie,
SMART, Mali https://www.unicef.org/mali/media/2321/file/SMART%202017.pdf

Wooldridge, J. M. (2007). Inverse probability weighted estimation for general missing data problems. *Journal of Econometrics*, *141*(2), 1281-1301.
28

Table 1: Descriptive Statistics

|  | Did not exclusively breastfed in the first six months | Exclusively breastfed in the first six months | Total sample |
|---|---|---|---|
| Majority exclusively breastfed in the first 6 months (Empirical exp.) | 0.147 | 0.553 | 0.225 |
| Majority approve of exclusive breastfeeding in the first 6 months (Normative exp.) | 0.241 | 0.553 | 0.301 |
| From Sikasso | 0.454 | 0.596 | 0.482 |
| Age (in years) (Mean) | 26.043 | 27.521 | 26.328 |
| Polygamous marriage | 0.290 | 0.362 | 0.304 |
| Literate | 0.364 | 0.266 | 0.345 |
| Proper living arrangements | 0.264 | 0.286 | 0.269 |
| Receive remittances from foreign | 0.183 | 0.213 | 0.189 |
| Participated in NGO programmes | 0.108 | 0.106 | 0.108 |
| Ethnicity: Bambara | 0.242 | 0.239 | 0.242 |
| Ethnicity: Fulas | 0.218 | 0.112 | 0.198 |
| Ethnicity: Sonikes | 0.155 | 0.133 | 0.151 |
| Ethnicity: Others | 0.385 | 0.516 | 0.410 |
| Observations | 788 | 188 | 976 |

*Source*: Survey data from Mali, 2019. For all the variables apart from age, the corresponding proportion in each group is given. The mean for age (in years) is given for all the groups.

Table 2: Marginal effects to estimate the association of empirical and normative expectations on breastfeeding behavior

|  | (1) | (2) | (3) | (4) | (5) |
|---|---|---|---|---|---|
| Majority exclusively breastfed in the first 6 months (Empirical exp.) | 0.228*** | 0.225*** | 0.184*** | 0.184*** | 0.222*** |
|  | (0.022) | (0.023) | (0.025) | (0.025) | (0.035) |
| Majority approve of exclusive breastfeeding in the first 6 months (Normative exp.) | 0.112*** | 0.116*** | 0.085*** | 0.087*** | 0.099*** |
|  | (0.023) | (0.024) | (0.025) | (0.025) | (0.031) |
| Age of the respondent |  | 0.002 | 0.002 | 0.002 | 0.001 |
|  |  | (0.002) | (0.002) | (0.002) | (0.002) |
| Polygamous marriage |  | 0.033 | 0.045* | 0.046** | 0.042 |
|  |  | (0.024) | (0.024) | (0.023) | (0.030) |
| Is the respondent educated |  | -0.046 | -0.033 | -0.038 | 0.004 |
|  |  | (0.029) | (0.027) | (0.027) | (0.035) |
| Proper living arrangements |  | 0.007 | 0.016 | 0.017 | 0.008 |
|  |  | (0.028) | (0.026) | (0.026) | (0.035) |
| Family received foreign remittance |  | 0.032 | 0.020 | 0.020 | 0.046 |
|  |  | (0.034) | (0.034) | (0.033) | (0.038) |
| Participated in NGO or Govt. development program |  | 0.025 | 0.046 | 0.048 | 0.056 |
|  |  | (0.041) | (0.038) | (0.038) | (0.043) |
| Ethnicity: Bambara |  |  |  | -0.023 | -0.037 |
|  |  |  |  | (0.040) | (0.058) |
| Ethnicity: Fulas/Peulh |  |  |  | -0.055 | -0.041 |
|  |  |  |  | (0.037) | (0.054) |
| Ethnicity: Soninkés |  |  |  | -0.038 | -0.178 |
|  |  |  |  | (0.050) | (0.111) |
| Sikasso |  | 0.068** | 0.163** | 0.125 | -0.022 |
|  |  | (0.031) | (0.066) | (0.077) | (0.092) |
| Cercle dummies |  | No | Yes | Yes | Yes |
| ES dummies |  | No | No | No | Yes |
| Pseudo R-squared | 0.152 | 0.178 | 0.218 | 0.221 | 0.283 |
| Observations | 976 | 921 | 921 | 921 | 757 |

*Notes:* The marginal effects from probit regression model are presented with standard errors, clustered at the ES level in the parentheses. Significance at *** $p < 0.01$, ** $p < 0.05$, * $p < 0.1$ level. The dependent variable is whether the mother exclusively breastfed her child or not in the first six months after birth.

Table 3: Comparison of basic variables across the different combinations of the vignettes

|  | Low EE-Low NE | High EE-Low EE | Low EE-High EE | High EE-High EE |
|---|---|---|---|---|
| Age (in years) | 26.242 | 26.168 | 26.397 | 26.593 |
| Polygamous marriage | 0.302 | 0.318 | 0.295 | 0.304 |
| Literate | 0.351 | 0.378 | 0.364 | 0.288 |
| Proper living arrangements | 0.254 | 0.261 | 0.260 | 0.304 |
| Receive remittances from foreign | 0.189 | 0.151 | 0.186 | 0.225 |
| Participated in NGO programmes | 0.106 | 0.126 | 0.103 | 0.093 |
| Ethnicity: Bambara | 0.238 | 0.265 | 0.227 | 0.242 |
| Ethnicity: Fulas | 0.208 | 0.181 | 0.169 | 0.233 |
| Ethnicity: Sonikes | 0.158 | 0.105 | 0.169 | 0.169 |
| Ethnicity: Others | 0.396 | 0.450 | 0.434 | 0.356 |
| From Sikasso | 0.430 | 0.563 | 0.467 | 0.470 |

*Source*: Survey data from Mali, 2019. For all the variables apart from age, the corresponding proportion in each group is given. The mean for age (in years) is given for all the groups.

Figure 1: Estimations for EE and NE bias

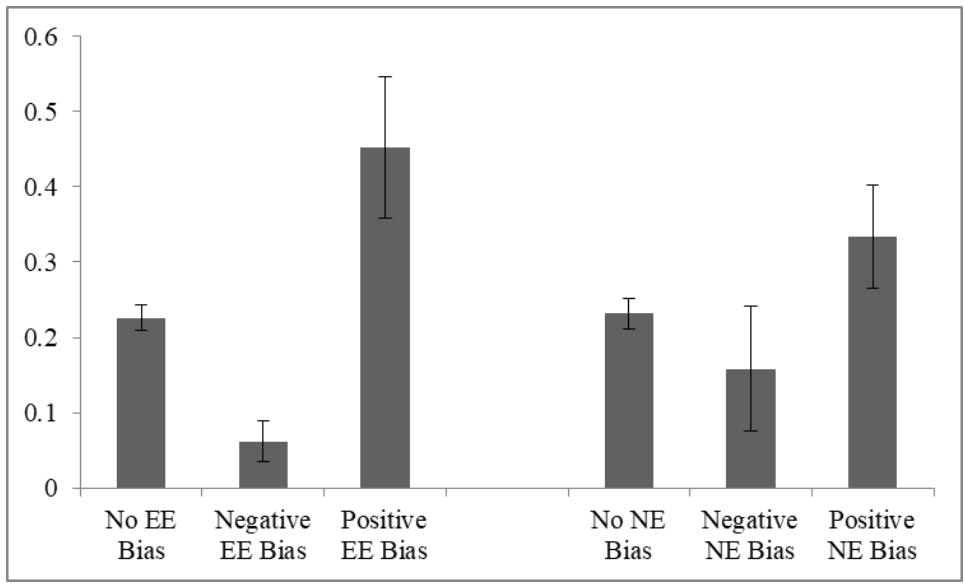

*Notes:* The marginal effects from probit regression model are presented with 95% confidence interval calculated by clustering the standard errors at the ES level. The dependent variable is whether the mother exclusively breastfed her child or not in the first six months after birth.

Figure 2: Prevalence of exclusive breast-feeding across the different combinations of the vignettes

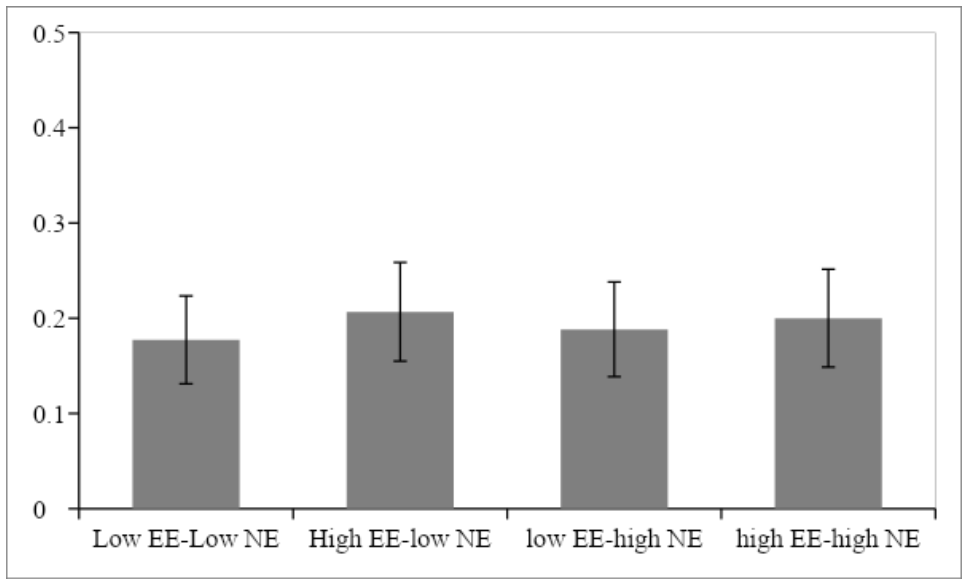

*Notes:* The marginal effects from probit regression model are presented with 95% confidence interval calculated by clustering the standard errors at the ES level. The dependent variable is whether the mother exclusively breastfed her child or not in the first six months after birth.

Figure 3: Regression estimates from randomly administered vignettes

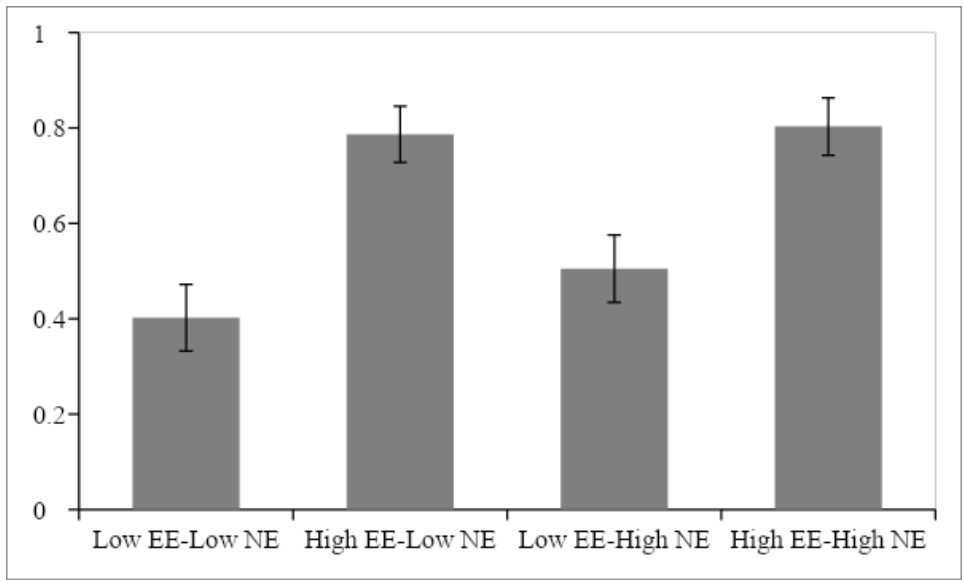

*Notes:* The marginal effects from probit regression model are presented with 95% confidence interval calculated by clustering the standard errors at the ES level. The dependent variable is whether the mother reported that the character in the vignette would exclusively breastfed her child or not in the first six months after birth.

Appendix A1: Marginal effects for EE and NE Bias

|  | (1) | (2) | (3) |
|---|---|---|---|
| *Ref. No EE bias* | | | |
| Negative EE bias | -0.175*** |  | -0.165*** |
|  | (0.014) |  | (0.018) |
| Positive EE bias | 0.248*** |  | 0.227*** |
|  | (0.053) |  | (0.056) |
| *Ref. No NE bias* | | | |
| Negative NE bias |  | -0.139*** | -0.074 |
|  |  | (0.033) | (0.049) |
| Positive NE bias |  | 0.142*** | 0.102** |
|  |  | (0.043) | (0.042) |
| Sikasso | 0.306*** | 0.308*** | 0.324*** |
|  | (0.091) | (0.104) | (0.095) |
| Ethnicity: Bambara | -0.036 | -0.036 | -0.038 |
|  | (0.056) | (0.062) | (0.058) |
| Ethnicity: Fulas/Peulh | -0.035 | -0.037 | -0.039 |
|  | (0.053) | (0.058) | (0.054) |
| Ethnicity: Soninkés | -0.155 | -0.155 | -0.172 |
|  | (0.112) | (0.119) | (0.110) |
| Age of the respondent | 0.002 | 0.001 | 0.001 |
|  | (0.002) | (0.002) | (0.002) |
| Polygamous marriage | 0.036 | 0.031 | 0.040 |
|  | (0.030) | (0.032) | (0.030) |
| Is the respondent educated | 0.002 | -0.003 | 0.004 |
|  | (0.035) | (0.035) | (0.035) |
| Proper living arrangements | 0.007 | 0.026 | 0.010 |
|  | (0.036) | (0.035) | (0.035) |
| Family received foreign remittance | 0.047 | 0.040 | 0.044 |
|  | (0.038) | (0.044) | (0.038) |
| Participated in NGO or Govt. development program | 0.062 | 0.061 | 0.055 |
|  | (0.043) | (0.047) | (0.042) |
| Cercle dummies | Yes | Yes | Yes |
| ES dummies | Yes | Yes | Yes |
| Pseudo R-square | 0.275 | 0.281 | 0.285 |
| Observations | 757 | 757 | 757 |

*Notes:* The marginal effects from probit regression model are presented with standard errors, clustered at the ES level in the parentheses. Significance at *** $p < 0.01$, ** $p < 0.05$, * $p < 0.1$ level. The dependent variable is whether the mother exclusively breastfed her child or not in the first six months after birth.

Appendix A2: Marginal effects for the experimental vignettes

|  | (1) | (2) | (3) | (4) | (5) |
|---|---|---|---|---|---|
| *Ref. Low EE-low NE* | | | | | |
| High EE-low NE | 0.404*** | 0.393*** | 0.374*** | 0.373*** | 0.397*** |
|  | (0.048) | (0.049) | (0.048) | (0.048) | (0.051) |
| Low EE-high NE | 0.111** | 0.105** | 0.089* | 0.088* | 0.113** |
|  | (0.046) | (0.046) | (0.048) | (0.047) | (0.053) |
| High EE-high NE | 0.404*** | 0.405*** | 0.396*** | 0.398*** | 0.432*** |
|  | (0.054) | (0.054) | (0.052) | (0.052) | (0.052) |
| Sikasso | | 0.041 | 0.263*** | 0.198* | -0.006 |
|  | | (0.043) | (0.096) | (0.111) | (0.093) |
| Bambara | | | | -0.021 | -0.014 |
|  | | | | (0.045) | (0.049) |
| Fulas/Peulh | | | | -0.026 | 0.044 |
|  | | | | (0.051) | (0.066) |
| Soninkés | | | | -0.083 | -0.095 |
|  | | | | (0.055) | (0.089) |
| age of the respondent | | -0.004 | -0.004 | -0.004 | -0.002 |
|  | | (0.003) | (0.003) | (0.003) | (0.003) |
| Mariage polygame | | 0.060 | 0.049 | 0.053 | 0.016 |
|  | | (0.043) | (0.041) | (0.040) | (0.042) |
| Is the respondent educated | | 0.079** | 0.084** | 0.077** | 0.088** |
|  | | (0.035) | (0.038) | (0.038) | (0.039) |
| Proper living arrangements | | 0.044 | 0.059 | 0.057 | -0.006 |
|  | | (0.041) | (0.042) | (0.041) | (0.041) |
| Family received foreign remittance | | 0.053 | 0.045 | 0.048 | 0.087* |
|  | | (0.048) | (0.047) | (0.047) | (0.049) |
| Participated in NGO or Govt. development program | | -0.017 | -0.000 | -0.000 | -0.005 |
|  | | (0.051) | (0.052) | (0.052) | (0.060) |
| Cercle dummies | No | No | Yes | No | Yes |
| ES dummies | No | No | No | No | Yes |
| Pseudo R-square | 0.106 | 0.112 | 0.146 | 0.149 | 0.246 |
| Observations | 954 | 896 | 896 | 896 | 840 |

*Notes:* The marginal effects from probit regression model are presented with standard errors, clustered at the ES level in the parentheses. Significance at *** $p < 0.01$, ** $p < 0.05$, * $p < 0.1$ level. The dependent variable is whether the hypothetical vignette character (Mariam) would exclusively breastfeed her child.